\documentclass[a4paper,12pt]{article}

\usepackage[margin=1in]{geometry}
\usepackage{setspace}
\usepackage[utf8]{inputenc}

\usepackage{amsmath}
\usepackage{amssymb}

\usepackage{graphicx}
\usepackage{float}
\usepackage[colorlinks,linkcolor=blue]{hyperref}
\usepackage{caption}
\usepackage{cite}
\usepackage{color}

\graphicspath{{Pictures/}}

\usepackage[affil-it]{authblk}

\title{\bf{Manifestation of proton structure in the initial-state anisotropies in high-energy proton-proton collisions}}

\author{Patryk Kubiczek\thanks{\texttt{patryk.kubiczek@student.uj.edu.pl}}}
\affil{Faculty of Physics, Astronomy and Applied Computer Science,\\ Jagiellonian University,\\
Łojasiewicza 11, 30-348 Kraków, Poland}

\author{Stanisław D. Głazek}
\affil{Institute of Theoretical Physics, Faculty of Physics, \\
University of Warsaw,
\\ Pasteura 5, 02-093 Warszawa, Poland}

\date{}

\begin{document}

\maketitle
\begin{abstract}
Ridge-like correlations in high-energy proton-proton 
collisions reported by the CMS collaboration suggest 
a collective flow that resembles the one in heavy-ion 
collisions. If the hydrodynamic description is valid 
then the effect results from the initial anisotropy 
of the colliding matter which depends on the 
structure of protons. Following recent theoretical 
developments, we propose several phenomenological 
models of the proton structure and calculate the 
anisotropy coefficients using the Monte Carlo Glauber 
model. Our estimates suggest that the event multiplicity 
dependence 
allows one to discriminate between different proton models.
\end{abstract}



%%%%%%%%%%%%%%%%%%%%%%
\section{Introduction}
%%%%%%%%%%%%%%%%%%%%%%

The analysis of two-particle angular correlations in 
\textit{pp} collisions at $\sqrt{s} = 7 \, \mathrm{TeV}$ 
revealed an unexpected near side ($\Delta \phi = 0$) 
correlation in the azimuthal angle of produced particles 
('ridge effect')~\cite{CMS}. 
There is no obvious reason why such a~long-range in pseudorapidity 
correlation should occur~\cite{bialkowska,li}. Ridge effect has 
been previously observed in relativistic heavy-ion collisions 
and it was explained by a collective anisotropic flow of hot and dense 
medium (hadronic gas or quark gluon plasma) created at the collision. 

It is possible that the
hydrodynamic explanation applies to high-multiplicity \textit{pp} 
events which are sufficiently energetic for the collision products 
to thermalize~\cite{bozek}. In this paper we assume 
validity of such a scenario and postulate that 
the proton internal structure can manifest itself in the ridge effect. 
During the hydrodynamic expansion, the spatial anisotropies of the 
collision area are transformed into anisotropic flow of produced 
particles. Thus, the spatial anisotropies 
due to the
proton structure 
should be reflected in the  
measurable
collective flow coefficients.

Actually, there exist in literature
estimates of the possible elliptic flow in \textit{pp} 
collisions~\cite{dent,casal,prasad,bozek2,ortona,pierog,bautista,
avsar,deng}. We basically follow the approach of~\cite{dent} in 
order to make a link between the 
proton
models considered by us and the experimental data. Within the 
framework of Glauber model~\cite{miller}, we calculate for each 
model the 
initial
spatial anisotropies: eccentricity, denoted by $\epsilon_2$, 
and triangularity, denoted by $\epsilon_3$, in proton-proton 
collision taking into account event-by-event fluctuations in 
the proton density profile.

The considered phenomenological proton densities are 
inspired by the quark-diquark~\cite{bbg} and RGPEP~\cite{glazek} 
models of proton. We find that it is possible to distinguish highly 
eccentric (rod-like) and highly triangular (triangle-like) fixed 
proton densities by looking at  
distributions of $\epsilon_2$ and $\epsilon_3$ and
the multiplicity dependence of their RMS values.
We also calculate these quantities for the
fluctuating anisotropic proton configurations 
(Gaussian-fluctuating model). 

We do not explicitly present results for the anisotropic 
flow coefficients $v_n$ because we limit our discussion 
to comparison of predictions of different models for 
the initial states in hydrodynamic 
evolution and the prediction of the exact shapes of the 
resulting final flow anisotropies 
is beyond the scope of this work. Our calculations 
concern the initial spatial anisotropies, though we emphasize 
that they should correspond to the flow coefficients measurable 
in multi-particle correlations in a fashion that is 
unlikely to be
far from linear \cite{v2, v3}. 

%%%%%%%%%%%%%%%%%%%%%%%%%%%%%%%%%%%%%%%%%%%%%%%%%%
\section{Glauber model for \textit{pp} collisions}
%%%%%%%%%%%%%%%%%%%%%%%%%%%%%%%%%%%%%%%%%%%%%%%%%%

The optical limit of Glauber model treats the collision of 
two composite particles as a superposition of independent 
binary collisions between their constituents. In case of 
\textit{pp} collisions, we assume the constituents are 
partons of one type,
and their smoothed distribution is 
given by proton density profile. We fix the partonic cross 
section $\sigma_{gg} = 4.3 \, \mathrm{mb}$ \cite{v2} and adjust
the number of partons in proton $N_g$ so 
that the experimental inelastic \textit{pp} cross 
section $\sigma_{pp} \approx 60 \, \mathrm{mb}$ \cite{cms1} is reproduced.

The geometrical quantities calculated within the
Glauber formalism correspond to the impact plane of the 
collision ($xy$) such that any $z$-dependence is integrated 
out. The density of binary collisions is given by the expression:    
\begin{equation}
n_{\mathrm{coll}}(x,y;b,\mathbf{\Sigma}_A, \mathbf{\Sigma}_B) 
= \sigma_{gg} \int_{-\infty}^{\infty} dz \, 
\rho\left(x - \frac{b}{2},y,z; \mathbf{\Sigma}_A\right) 
\int_{-\infty}^{\infty} dz' \, \rho\left(x + \frac{b}{2},y,z'; \mathbf{\Sigma}_B\right),
\end{equation}
where $\rho(x,y,z; \mathbf{\Sigma})$ is a proton density 
profile, $b$ is an impact parameter and by $\mathbf{\Sigma}$ 
we mean a set of fluctuating parameters describing the proton 
internal structure and orientation in space (for fixed 
configuration models $\mathbf{\Sigma}$ dependence will 
correspond only to possible rotations in space) \cite{miller}.

For each event characterized by $b$, $\mathbf{\Sigma}_A$ and $\mathbf{\Sigma}_B$ we calculate eccentricity $\epsilon_2$ and 
triangularity $\epsilon_3$. We choose the following definition 
of spatial anisotropy \cite{blaizot}:
\begin{equation}
\epsilon_n = \frac{\sqrt{\left\{ s^n \cos(n \phi)\right\}^2 
+\left\{ s^n \sin(n \phi)\right\}^2}}{\left\{ s^n \right\}},
\end{equation}
where $\phi$ is an azimuthal angle in $xy$ plane and $s^2 = 
x^2 + y^2$. The curly brackets $\left\{ \ldots \right\}$ 
stand for the average taken with respect to the impact 
plane binary collision density $n_{\mathrm{coll}} (x,y;b,\mathbf{\Sigma}_A, 
\mathbf{\Sigma}_B)$:
\begin{equation}
\left\{ f(x,y) \right\} = \frac{\int dx \, dy \, f(x, y) n_{\mathrm{coll}}(x,y;b,\mathbf{\Sigma}_A, 
\mathbf{\Sigma}_B)}{\int dx \, dy \, n_{\mathrm{coll}}(x,y;b,\mathbf{\Sigma}_A, 
\mathbf{\Sigma}_B)}
\end{equation}

The differential cross section (probability) of a given event 
depends only on the integrated number of binary collisions 
$N_{\mathrm{coll}}(b,\mathbf{\Sigma}_A, \mathbf{\Sigma}_B) 
= \int dx \, dy \, n_{\mathrm{coll}}(x,y;b,\mathbf{\Sigma}_A, 
\mathbf{\Sigma}_B)$,
\begin{equation}
\sigma(b,\mathbf{\Sigma}_A, \mathbf{\Sigma}_B) =
1 - \left[ 1 - \frac{N_{\mathrm{coll}}(b,\mathbf{\Sigma}_A, 
\mathbf{\Sigma}_B)}{N_g^2}\right]^{N_g^2}.
\label{eq:cross-section} 
\end{equation}
The expectation value of a quantity $Q$ is calculated in the following way
\begin{equation}
\left< Q \right> = \frac{1}{\sigma_{pp}} \int_0^{\infty} 2\pi 
b \, db \int P(\mathbf{\Sigma}_A) \, d \mathbf{\Sigma}_A 
\int P(\mathbf{\Sigma}_B) \, d \mathbf{\Sigma}_B \, 
\sigma(b,\mathbf{\Sigma}_A, \mathbf{\Sigma}_B) \, 
Q(b,\mathbf{\Sigma}_A, \mathbf{\Sigma}_B),
\label{eq:mean} 
\end{equation}
where $\sigma_{pp} =  \int_0^{\infty} 2\pi b \, db \int 
P(\mathbf{\Sigma}_A) \, d \mathbf{\Sigma}_A \int 
P(\mathbf{\Sigma}_B) \, d \mathbf{\Sigma}_B \, 
\sigma(b,\mathbf{\Sigma}_A, \mathbf{\Sigma}_B)$ 
and $P(\mathbf{\Sigma})$ is the probability density 
of proton configuration $\mathbf{\Sigma}$.

We calculate charged hadron multiplicity $N$ for 
each event assuming its linear scaling with
$N_\mathrm{coll}$:
\begin{equation}
N = \alpha N_\mathrm{coll}, 
\label{eq:multiplicity}
\end{equation}
where $\alpha$ is determined by the demand of reproducing 
experimental minimum bias charged hadron multiplicity 
$\left< N \right> = 30$ \cite{cms2}. 

There is another approach to estimating the multiplicities 
and geometrical quantities within the Glauber formalism, 
namely the wounded nucleon model \cite{bialas}. Within this 
model, all the averages denoted by $\left\{ \ldots \right\}$ are taken 
with respect to the local density of participating constituents 
instead of the density of binary collisions. However, in high 
energy hadronic collisions, multiple gluonic interactions are 
expected to occur, so we think it is more natural to use here 
the density of binary partonic collisions.

%%%%%%%%%%%%%%%%%%%%%%%%%%%%%%%%%%%%%%%%%%%%%
\section{Models of proton internal structure}
%%%%%%%%%%%%%%%%%%%%%%%%%%%%%%%%%%%%%%%%%%%%%

The renormalization group procedure for effective particles 
(RGPEP)~\cite{glazek} suggests that a proton can be described 
in terms of three effective quarks and a gluon body interacting 
via harmonic potential. For every considered special case of 
this picture we model the proton as a sum of isotropic 3D 
Gaussian densities $\rho_q$ corresponding to effective quarks 
and anisotropic Gaussians $\rho_g$ representing the gluonic 
flux tubes connecting the quarks:
\begin{eqnarray}
\rho_q(\mathbf{r}; r_q) & = & \frac{1}{(2 \pi)^{3/2} \, 
r_q^3} e^{-\frac{r^2}{2 r_q^2}} \ , \\
\rho_g(\mathbf{r}; r_s, r_l) & = & 
\frac{1}{(2 \pi)^{3/2} \, r_s^2 \, r_l} 
e^{-\frac{x^2 + y^2}{2 r_s^2} -\frac{z^2}{2
r_l^2}} \ .
\end{eqnarray}
The effective quarks and the gluon body are clusters of 
partons, as in the 
two stage and cluster models \cite{altarelli, hove, hwa} 
used to explain the shape of deep inelastic scattering 
structure functions.

%%%%%%%%%%%%%%%%%%%%%%%%%%%%%%%%%%%%%%%%%%%%%%%%
\subsection{Fixed configuration models: I and Y}
%%%%%%%%%%%%%%%%%%%%%%%%%%%%%%%%%%%%%%%%%%%%%%%%

According to \cite{bbg}, the ridge effect 
could be the 
consequence of large eccentricities $\epsilon_2$ in 
high-multiplicity collisions between aligned quark-diquark 
states of protons which constitute around 20\% of all 
possible states.

\begin{figure}
\centering
\includegraphics[width=0.60\linewidth]{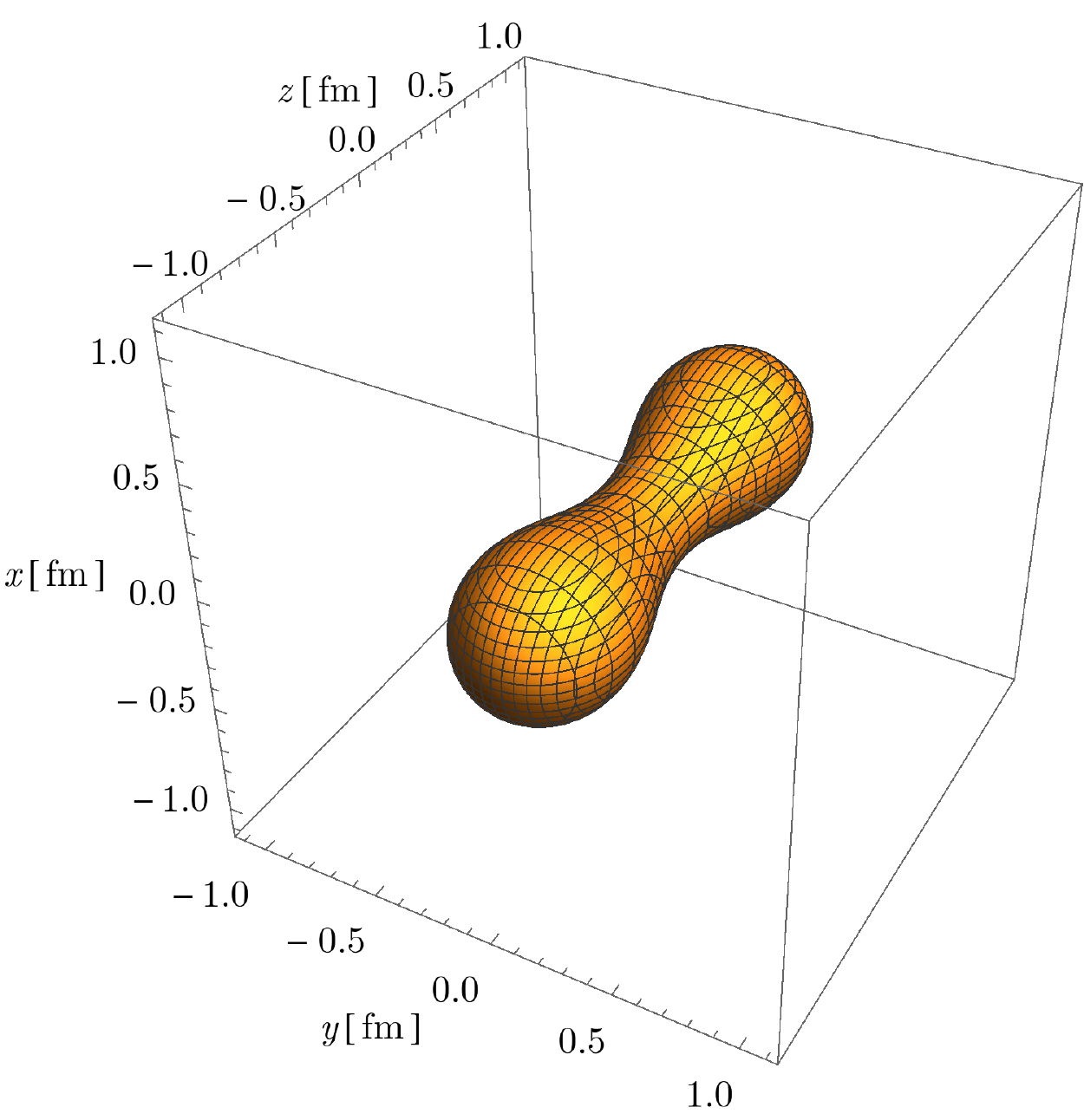}
\caption{Constant proton density surface for I (quark-diquark) model}  
\label{f:density-I}
\end{figure}

We model this state, 
labelling it with the symbol I, 
as a superposition of two effective 
quark bodies, with quark and diquark 
parton densities generically described 
in terms of simple functions, 
and two 
gluonic tubes connecting the quark bodies
(Fig. \ref{f:density-I}).
Such greatly simplified density model
is sufficient for tracing the effects 
of symmetric rod-like shape of proton
in comparison with other conceivable shapes
(see below), leaving the more subtle effects 
of the asymmetry between quark and diquark for further analysis elsewhere.
Thus, we use here
\begin{align}
\rho_{\mathrm{I}}(\mathbf{r}) & = N_g  
\frac{1-\kappa}{2} \left[ \rho_q\left(x, y, z - \frac{d}{2}; r_q\right) 
+ \rho_q\left(x, y, z + \frac{d}{2}; r_q\right) \right] \nonumber \\
 & + N_g \frac{\kappa}{2} \left[ \rho_g\left(x, y, z - \frac{d}{4}; r_q, 
\frac{d}{4}\right) + \rho_g\left(x, y, z + \frac{d}{4}; r_q, \frac{d}{4}\right) \right] 
\ .
\end{align}
The free parameters of the model are: 
the effective quark radius $r_q$,  
the length of quark-diquark tube $d$, 
and the percentage of gluon body content 
$\kappa$. Assuming that only effective 
quark bodies carry
net charge 
($+4/3 e$ homogenously distributed in the diquark and $-1/3 e$ in the quark), 
we choose $r_q = 0.25 \, 
\mathrm{fm}$ and $d = 1.5  \, \mathrm{fm}$, 
which reproduce charge rms radius of proton 
($\approx 0.9 \, \mathrm{fm}$). 
In this work, we always choose $\kappa = 0.5$.   
 
\begin{figure}
\centering
\includegraphics[width=0.60\linewidth]{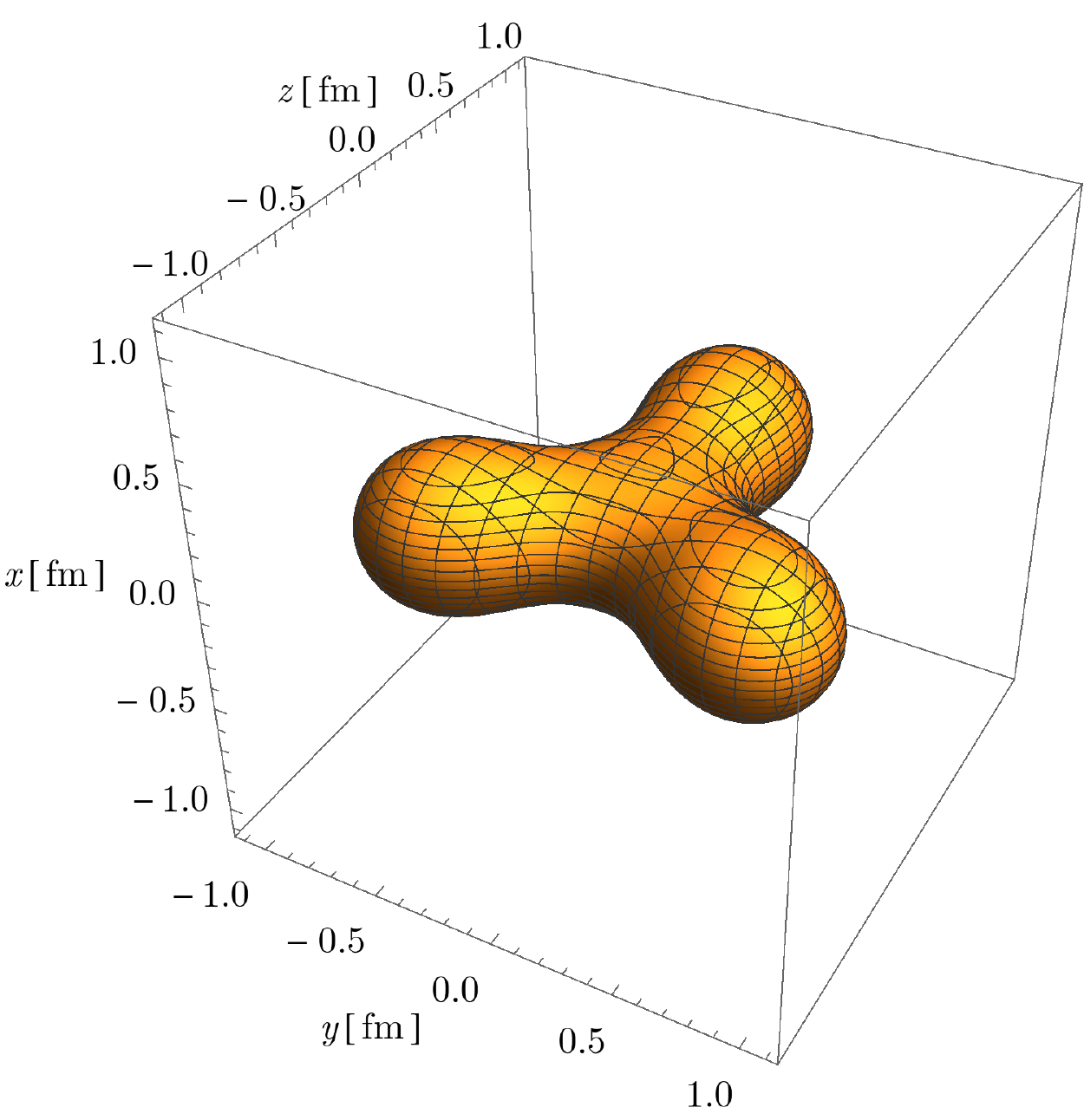}
\caption{Constant proton density surface for Y model}  
\label{f:density-Y}
\end{figure}

For comparison, we also 
consider a model that we label by the symbol Y
in which proton has a highly triangular shape (Fig. 
\ref{f:density-Y}). The 
parton
density in this case is assumed in the form
\begin{align}
\rho_{\mathrm{Y}}(\mathbf{r})  = N_g  \frac{1-\kappa}{3} 
\sum_{k=1}^{3} \rho_q(\mathbf{r - r}_k; r_q)  
  + N_g \frac{\kappa}{3} \sum_{k=1}^{3} 
\rho_g\left(\mathcal{R}^{-1}[k \frac{2\pi}{3}](\mathbf{r} 
- \frac{\mathbf{r}_k}{2}); r_q, \frac{r_k}{2}\right), 
\end{align}
where $\mathbf{r}_1 = \left(0, \frac{\sqrt{3}}{4} d, 
- \frac{d}{4} \right), \mathbf{r}_2 = \left(0, -\frac{\sqrt{3}}{4} d, - \frac{d}{4} \right), 
\mathbf{r}_3 = \left(0, 0,  \frac{d}{2} \right)$ and 
$\mathcal{R}[\vartheta]$ is a rotation matrix in $yz$ plane. 
In this model, the effective quarks are located in the 
vertices of an equilateral triangle and the gluon tubes 
connect them with the center of mass of the system. 
Again we choose $r_q = 0.25 \, \mathrm{fm}$ and $d = 
1.5  \, \mathrm{fm}$.

The densities of types
I and Y are fixed proton structure models in a sense 
that the only parameters that fluctuate event-by-event 
are the angles describing proton orientation relative 
to the direction of its velocity (not explicitly 
introduced above). 

%%%%%%%%%%%%%%%%%%%%%%%%%%%%%%%%%%%%%%%%%%%%%%%%%%%%%
\subsection{Gaussian-fluctuating configuration model}
%%%%%%%%%%%%%%%%%%%%%%%%%%%%%%%%%%%%%%%%%%%%%%%%%%%%%

We also consider a model with a
fluctuating proton configuration, labelled G-f, in which 
the relative positions of effective quarks differ event-by-event. 
By generalizing the previous formulas for proton densities, 
we introduce Gaussian-fluctuating proton density
\begin{align}
\rho_{\mathrm{G-f}}(\mathbf{r}; \mathbf{r}_1, \mathbf{r}_2, \mathbf{r}_3)  
= N_g  \frac{1-\kappa}{3} \sum_{k=1}^{3} \rho_q(\mathbf{r - r}_k; r_q)  
  + N_g \frac{\kappa}{3} \sum_{k=1}^{3} 
\rho_g\left[\mathcal{R}^{-1}[\theta_k, \phi_k](\mathbf{r} - 
\frac{\mathbf{r}_k}{2}); r_q, \frac{r_k}{2}\right]
.
  \end{align}
In this expression, $\mathcal{R}[\theta, \phi]$ transforms 
vector $(0,0,1)$ into $(\cos \phi \sin \theta, \sin \phi \sin 
\theta, \cos \theta)$ and $\mathbf{r}_k = r_k (\cos \phi_k 
\sin \theta_k, \sin \phi_k \sin \theta_k, \cos \theta_k)$ 
is the position vector of $k$-th effective quark.

Following the harmonic oscillator phenomenology, we assume 
that the probability density of finding a proton whose quarks 
are in positions $\mathbf{r}_1, \mathbf{r}_2, \mathbf{r}_3$ 
is Gaussian with an additional constraint of 
in the center-of-mass frame ($\mathbf{r}_1 + \mathbf{r}_2 + \mathbf{r}_3 = 0$, obtained by recentering of the quarks),
\begin{equation}
P(\mathbf{r}_1, \mathbf{r}_2, \mathbf{r}_3) 
= \frac{1}{[(2 \pi)^{3/2} \, R_P ^ 3]^3} 
e^{-(r_1^2 + r_2^2 + r_3^2) / 2 R_P^2} 
\ .
\end{equation}
The free parameters 
in the G-f model
are $\kappa$, $r_q$ and $R_P$. 
Again
we set 
$r_q = 0.25 \, \mathrm{fm}$, which implies the value 
$R_P = 0.43 \, \mathrm{fm}$, if one wants to obtain 
the experimental rms proton charge radius.

%%%%%%%%%%%%%%%%% 
\section{Results}
%%%%%%%%%%%%%%%%%

\begin{figure}
\centering
\includegraphics[width=0.80\linewidth]{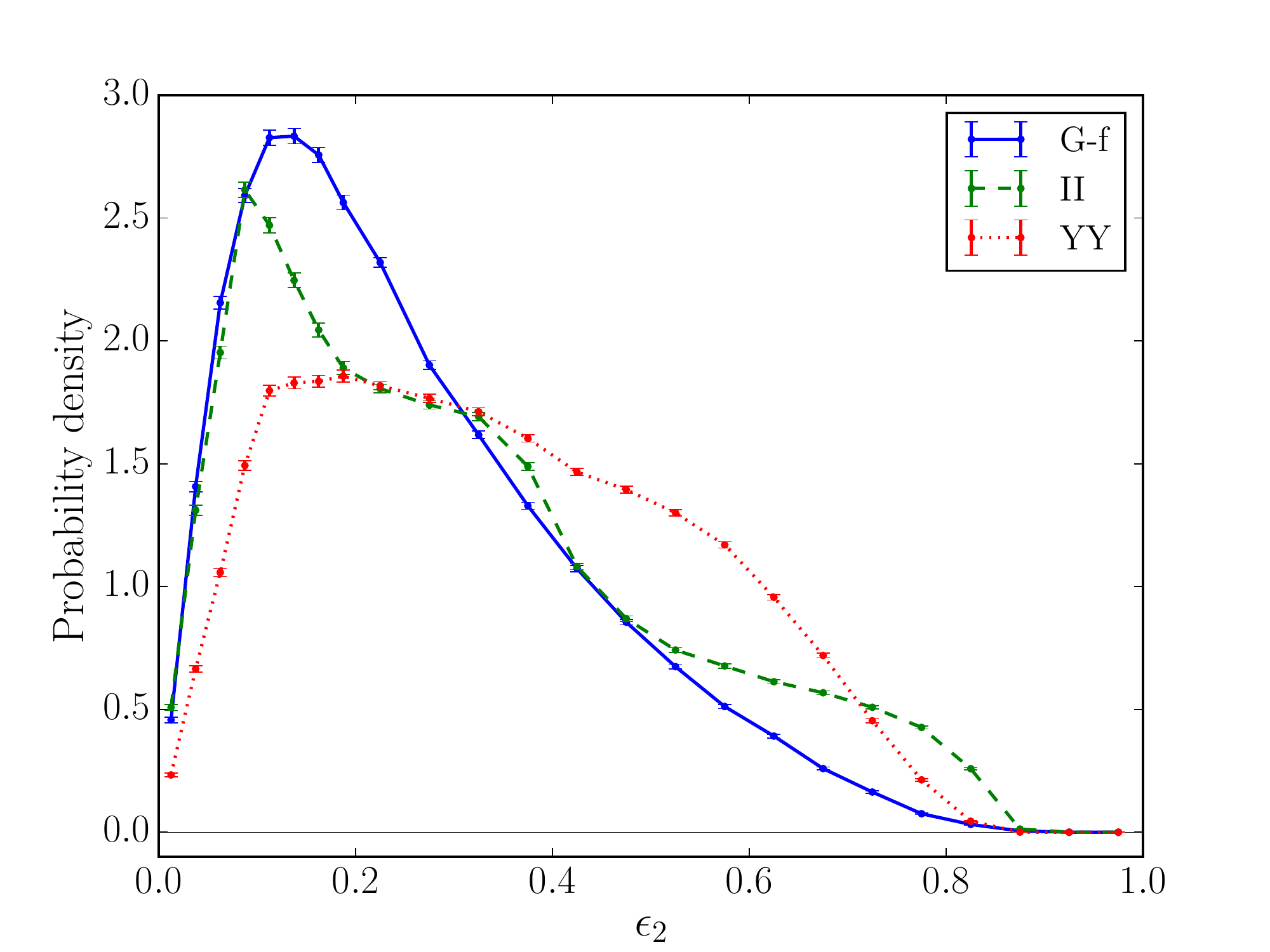}
\caption{Distribution of event eccentricities for II, YY and Gaussian-fluctuating type (G-f\,G-f) collisions. The continuous line is reconstructed from the values in particular eccentricity bins and represents the normalized probability density of the occurence of an event with certain eccentricity. Errors (square roots of variances) follow from basic statistics assuming independent events (applies to all other figures). }
\label{f:ecc}
\end{figure}

\begin{figure}
\centering
\includegraphics[width=0.80\linewidth]{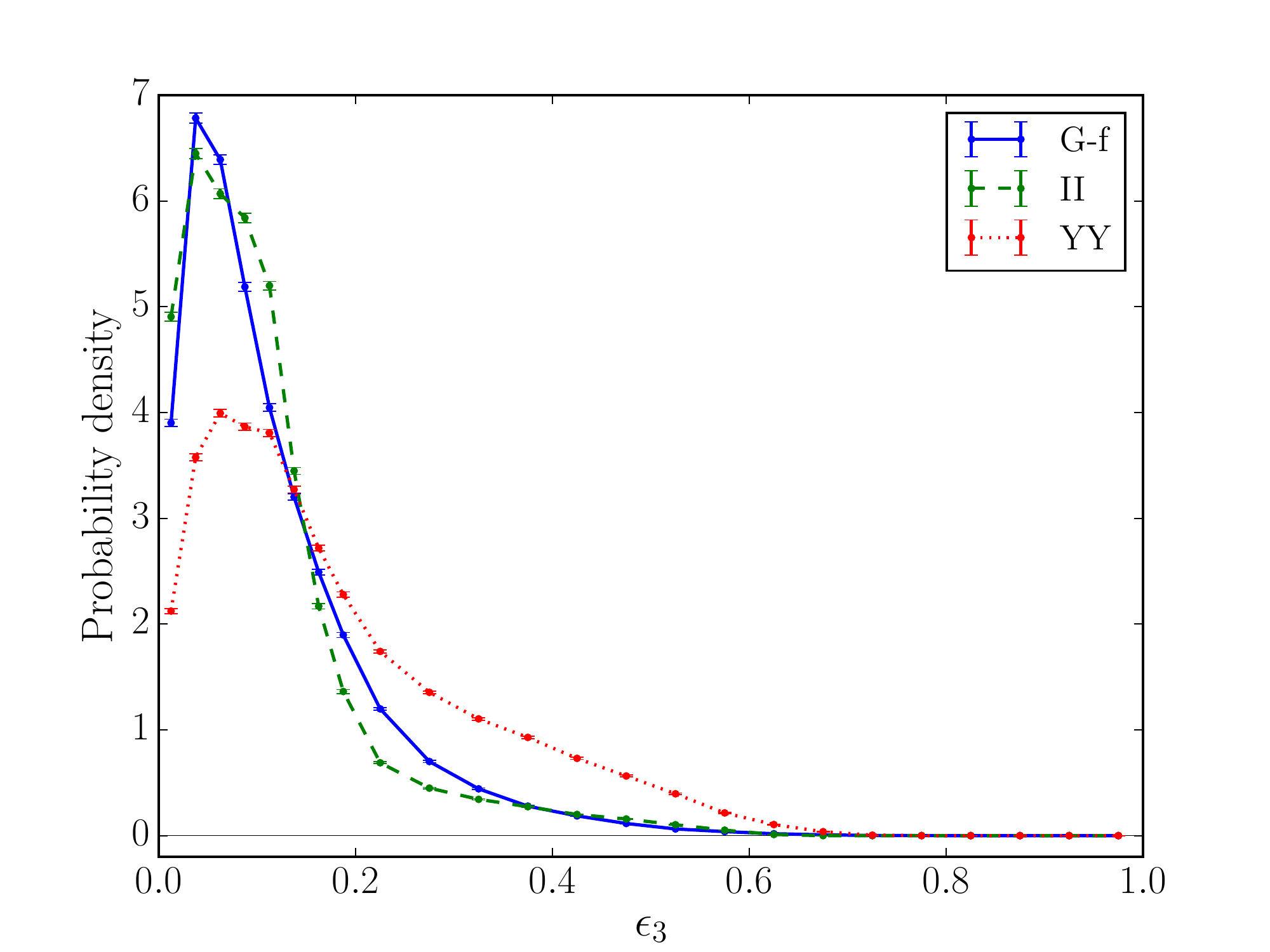}
\caption{Distribution of event triangularities for II, YY and Gaussian-fluctuating type (G-f\,G-f) collisions}
\label{f:tri}
\end{figure}

\begin{figure}
\centering
\includegraphics[width=0.80\linewidth]{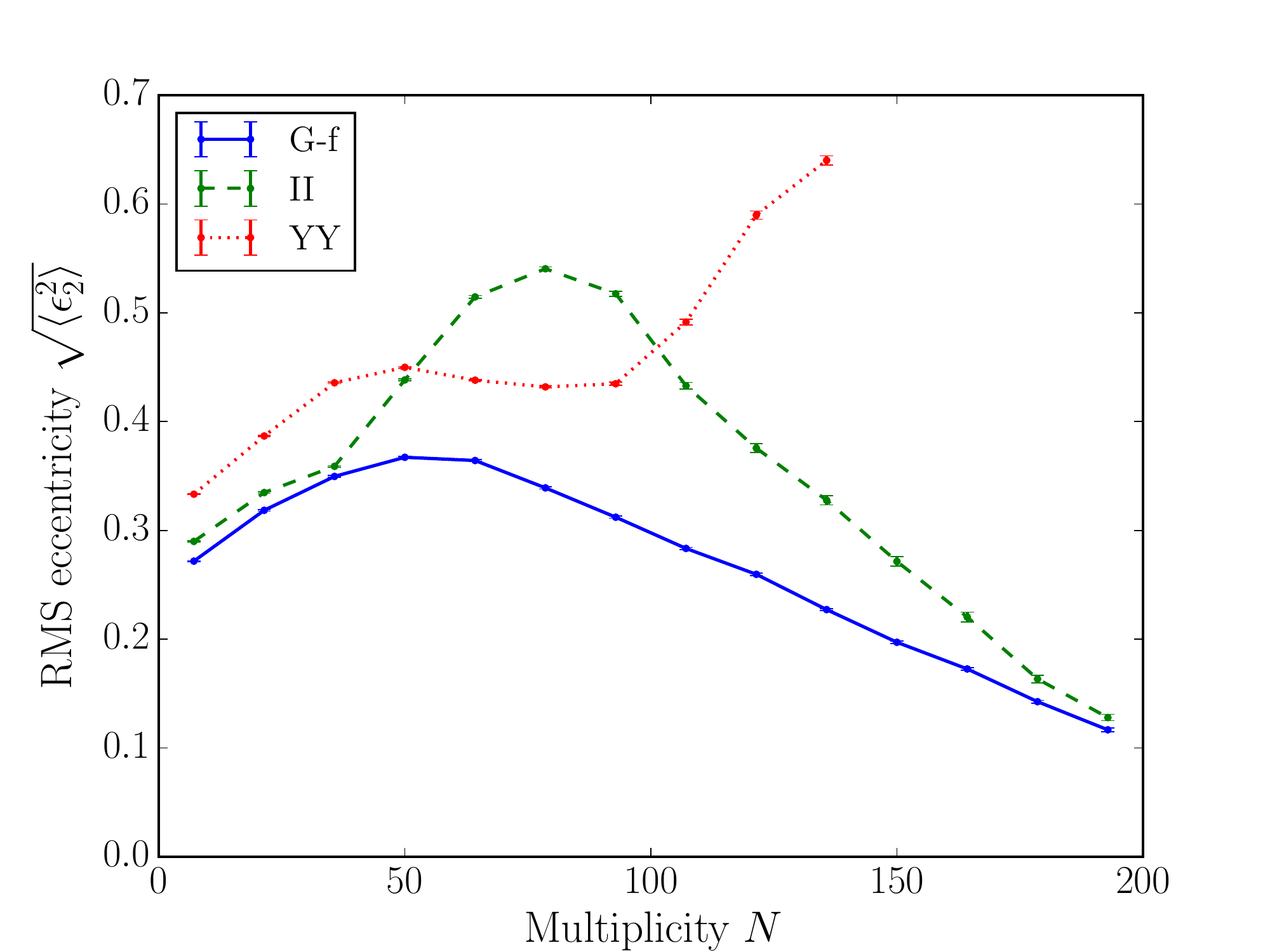}
\caption{RMS eccentricities in collision multiplicity bins for II, YY and Gaussian-fluctuating type (G-f\,G-f) collisions}
\label{f:mean-ecc}
\end{figure}

\begin{figure}
\centering
\includegraphics[width=0.80\linewidth]{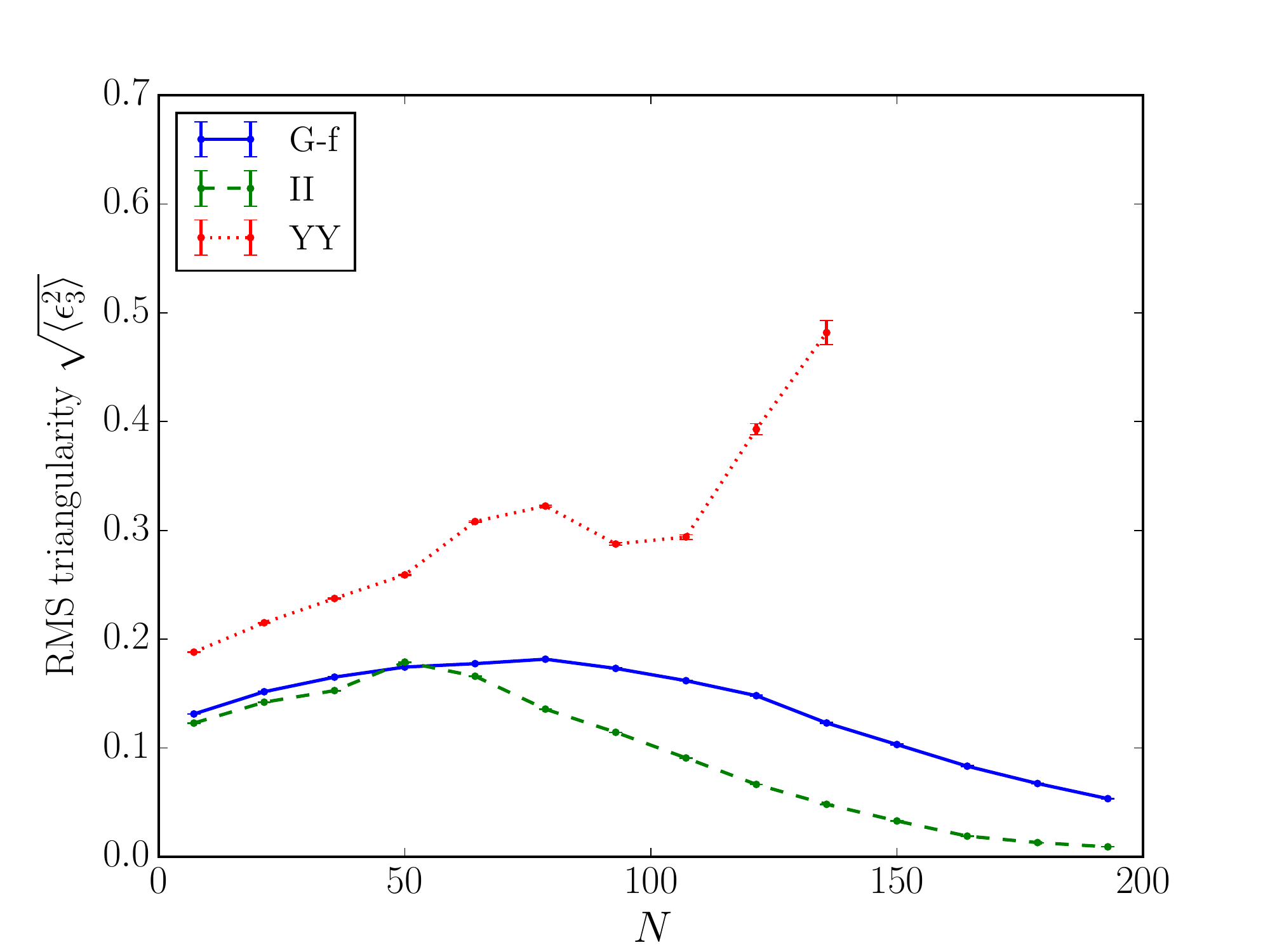}
\caption{RMS triangularities in collision multiplicity bins for II, YY and Gaussian-fluctuating type (G-f\,G-f) collisions}
\label{f:mean-tri}
\end{figure}

\begin{figure}
\centering
\includegraphics[width=0.80\linewidth]{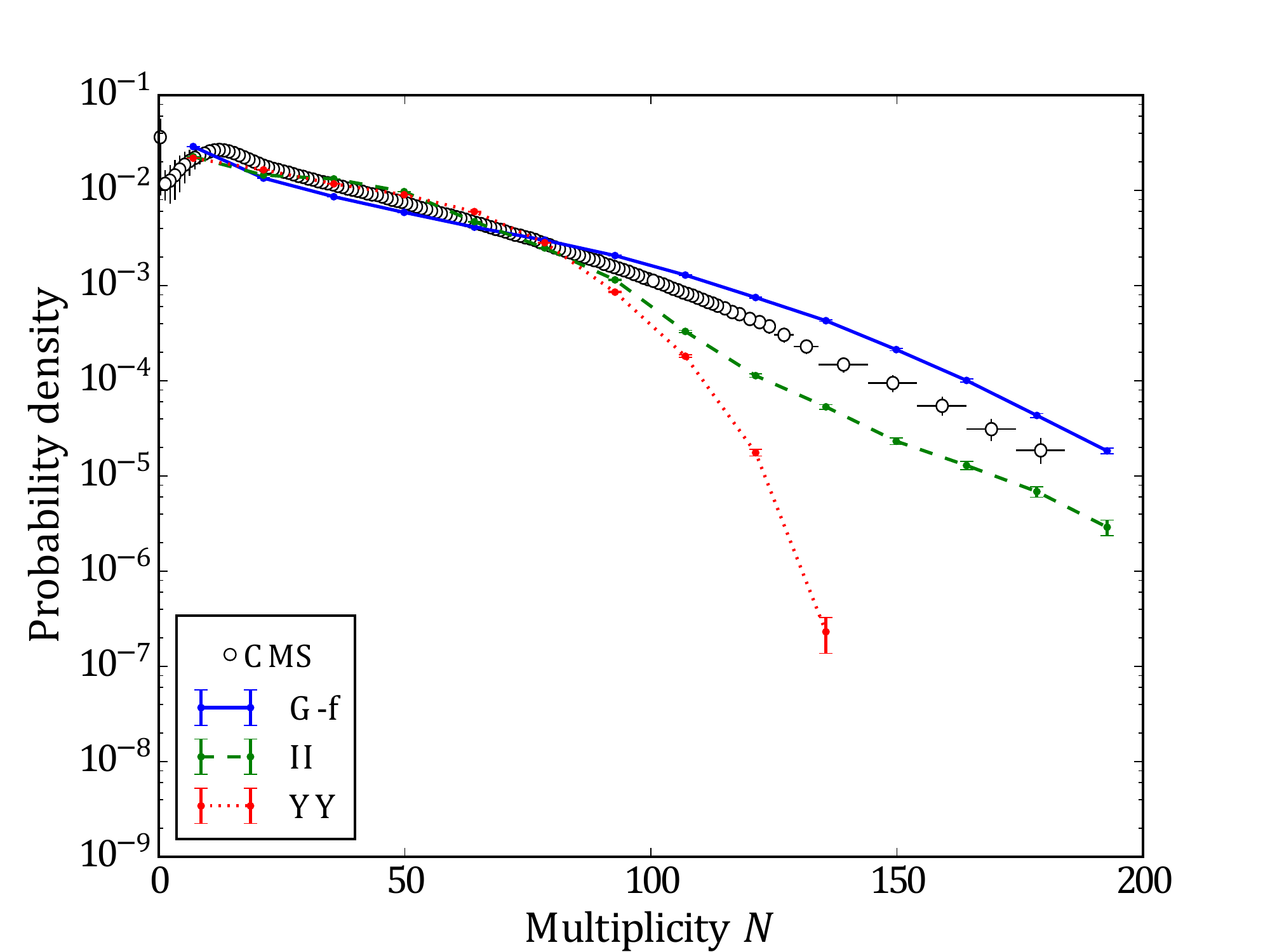}
\caption{Event multiplicity distributions for II, YY and Gaussian-fluctuating type collisions compared to the experimental charged hadron distribution in $\sqrt{s} = 7 \, \mathrm{TeV}$ \textit{pp} collisions \cite{cms2} for the pseudorapidity range $|\eta| < 2.4$}
\label{f:multi}
\end{figure}

We present the event distribution of eccentricities 
$\epsilon_2$ and triangularities $\epsilon_3$ for three 
classes of \textit{pp}
collision models: II, YY and G-f in which colliding 
protons are described by I, Y or G-f
model, respectively (Fig. \ref{f:ecc}, \ref{f:tri}). 
We also present the RMS values $\sqrt{\left< \epsilon^2_2 \right>}$  
and $\sqrt{\left< \epsilon^2_3 \right>}$ in certain multiplicity 
bins (Fig. \ref{f:mean-ecc}, \ref{f:mean-tri}) and the 
distribution of event multiplicities $N$ (Fig. \ref{f:multi}). 
The reason for presenting RMS and not mean values of anisotropies 
is because the former correspond hydrodynamically to the actual 
anisotropic flow coefficients extracted from two particle 
correlations (we base this approach on \cite{avsar}).
The minimum bias (averaged over all events) $\sqrt{\left< 
\epsilon^2_2 \right>_{\mathrm{MB}}}$ are 0.37, 0.40 and 0.31 
for II, YY and G-f model, respectively, while $\sqrt{\left< 
\epsilon^2_3 \right>_{\mathrm{MB}}}$ = 0.14, 0.23, 0.15, correspondingly 
(these quantities can be calculated using probability 
densities from Fig. \ref{f:ecc} and Fig. \ref{f:tri}).

All the calculations are performed by means of a Monte Carlo 
algorithm. 300 000 events are generated for each class of 
collision. The statistical weight of an event is given by 
Eq. (\ref{eq:cross-section}) and the mean values are calculated 
according to Eq. (\ref{eq:mean}). In case of I and Y models, 
we average only over the impact parameter $b$ and over all 
the possible space orientations of protons during the collision. 
In case of G-f model, the 
additional averaging over the positions of 
effective quarks takes place. 
The event multiplicity $N$ is estimated by the use of 
Eq. (\ref{eq:multiplicity}).
 
We set $N_g$ in such a way that the experimental \textit{pp} 
inelastic cross section of 60 mb is reproduced. This is 
accomplished for $N_g$ in range 7-10. Similarly, the parameter 
$\alpha$ from Eq. (\ref{eq:multiplicity}) is found to  vary 
between 5 and 11.

%%%%%%%%%%%%%%%%%%%%%
\section{Conclusions}
%%%%%%%%%%%%%%%%%%%%%

Our calculations predict quite high values of eccentricities 
and triangularities in \textit{pp} collisions. Assuming 
the hydrodynamic limit in which the scaling between the initial spatial and finite momentum anisotropy is linear and taking $v_2 / \epsilon_2 \sim 0.3$ \cite{v2}, we expect minimum bias $v_2 \sim 0.11, 0.12, 0.09$ for quark-diquark, triangular and Gaussian-fluctuating model. Bo\.{z}ek extracts from the ridge effect $v_2$ in range of 0.04 - 0.08 \cite{bozek} which can potentially be in agreement with our results if viscous effects reducing the scaling factor are taken into account.

Our models predict 
that one in principle 
should
be able to distinguish 
quark-diquark (I), triangular (Y) and Gaussian-fluctuating 
(G-f) proton configurations by extracting the collective flow 
coefficients $v_2$ and $v_3$ from multi-particle correlations. 
The actual shapes of the collective flow dependence on 
multiplicity will differ from Fig. \ref{f:mean-ecc} and 
Fig. \ref{f:mean-tri}, though it is reasonable to assume 
that to a first approximation the scaling of $v_n$ with 
$\epsilon_n$ is linear.
In general, the anisotropies in collisions of triangular protons, 
denoted as YY, are much higher than in those of II or G-f\,G-f. 
This was expected because
irrespective of
the orientation of Y protons the collision area is 
always anisotropic. We find that the values of anisotropy 
coefficients in YY 
collisions 
increase strongly in high 
multiplicity bins $N$. 
Collisions of quark-diquark
protons
(II) are characterised by 
relatively large eccentricities $\epsilon_2$ (though not 
as large
as in YY collisions) and small triangularities 
$\epsilon_3$. The distributions of anisotropies and the 
$N$ dependence of their mean values are the most smooth 
for the G-f\,G-f collisions, which results 
from the existence of additional degrees of freedom. 

We observe that the fraction of high-multiplicity events 
in the fixed configuration models is smaller than in the 
Gaussian-fluctuating model.
Our comparison to the 
experimental data (Fig. \ref{f:multi}) indicates that a 
more realistic model of $\textit{pp}$ collision
should include not 
just one type, but all II, YY, G-f\,G-f 
(and IY, I\,G-f, Y\,G-f) types of  
collision configurations with certain probabilities. 
However, the G-f model alone can
approximately describe the experimental results 
for multiplicity distributions.

Avsar \textit{et al.} \cite{avsar} using DIPSY Monte Carlo 
generator predict values of eccentricity which are similar 
to our result for triangular proton configurations. The 
distributions of eccentricities within the 'hot spot' model 
\cite{casal} seem to be centered more towards higher values 
of $\epsilon_2$ than in our result. The probable reason is 
that the 'hot spot' model 
lacks any constraint on the relative positions of quarks. It is hard to 
make any direct comparison with the other mentioned calculations, however all of them \cite{dent,casal,prasad,bozek2,ortona,
pierog,bautista,avsar,deng} predict measurable anisotropies 
in \textit{pp} collisions.

%%%%%%%%%%%%%%%%%
\section{Summary}
%%%%%%%%%%%%%%%%%

In view of the great interest in understanding proton 
structure, it is pointed out that even simple model 
ideas concerning distribution of quarks and gluons in 
protons lead to non-trivial multiplicity dependence 
of initial anisotropies in \textit{pp} collisions.
We illustrate this statement 
with results obtained for phenomenological models intuitively inspired by RGPEP in QCD and similar ideas.
Expecting new data on high-energy \textit{pp} collisions 
from the $\sqrt{s} = 14\, \mathrm{TeV}$ runs
of the Large Hadron Collider, we believe that 
detailed studies of the role of proton structure 
models in description of such collisions may 
shed new light on the internal structure of protons.

\newpage

\listoffigures

\newpage

\end{document}